\documentclass[3p,times,twocolumn,natbib209]{elsarticle}

\usepackage{lineno,natbib}
\modulolinenumbers[5]

\newcommand{\nat}{Nature,\ }

\journal{Journal of High Energy Astrophysics}

\biboptions{authoryear}

\makeatletter
\renewcommand\@biblabel[1]{}
\makeatother

\begin{document}


\begin{frontmatter}

\title{The First Ten Years of Swift Supernovae }

\author{Peter J. Brown}
\address{George P. and Cynthia Woods Mitchell Institute for Fundamental Physics \& Astronomy, \\
Texas A. \& M. University, Department of Physics and Astronomy, \\
4242 TAMU, College Station, TX 77843, USA; 
pbrown@physics.tamu.edu}

\author{Peter W. A. Roming}
\address{Southwest Research Institute, Department of Space Science, \\
6220 Culebra Road, San Antonio, TX 78238, USA}

\author{Peter A. Milne}
\address{University of Arizona, Steward Observatory, \\
933 North Cherry Avenue, Tucson, AZ 85719, USA}




\begin{abstract}

The Swift Gamma Ray Burst Explorer has proven to be an incredible platform for studying the multiwavelength properties of supernova explosions.  In its first ten years, Swift has observed over three hundred supernovae.  The ultraviolet observations reveal a complex diversity of behavior across supernova types and classes.  Even amongst the standard candle type Ia supernovae, ultraviolet observations reveal distinct groups.  When the UVOT data is combined with higher redshift optical data, the relative populations of these groups appear to change with redshift.  Among core-collapse supernovae, Swift discovered the shock breakout of two supernovae and the Swift data show a diversity in the cooling phase of the shock breakout of supernovae discovered from the ground and promptly followed up with Swift.  Swift observations have resulted in an incredible dataset of UV and X-ray data for comparison with high-redshift supernova observations and theoretical models.  Swift's supernova program has the potential to dramatically improve our understanding of stellar life and death as well as the history of our universe.

\end{abstract}


\end{frontmatter}



\section{Swift as a Supernova Observatory}

Though designed as a gamma-ray burst (GRB) follow-up mission, the Swift satellite \citep{Gehrels_etal_2004} has proven to be remarkable observatory for a different type of cosmic explosion -- supernovae.  Its short term scheduling and ability to be rapidly repointed allow observations to begin shortly after a supernova is discovered.  The discoveries usually come from ground-based supernova searches, though Swift actually discovered two unique supernovae based on their shock breakout emission which are discussed below.  Swift's short term scheduling allows campaigns to be modified regularly based on the results of data which are available on the internet hours after the observations.  Viewing constraints require Swift to point to several different objects within its ninety minute orbit, so there is no increased overhead for short observations of relatively bright objects.  This allows nearby supernovae to be observed with a relatively short cadence to follow the temporal evolution.  These observing characteristics are coupled to the unique nature of ultraviolet (UV) and X-ray observations which cannot be made from the ground.

\begin{figure} 
\includegraphics[scale=0.5]{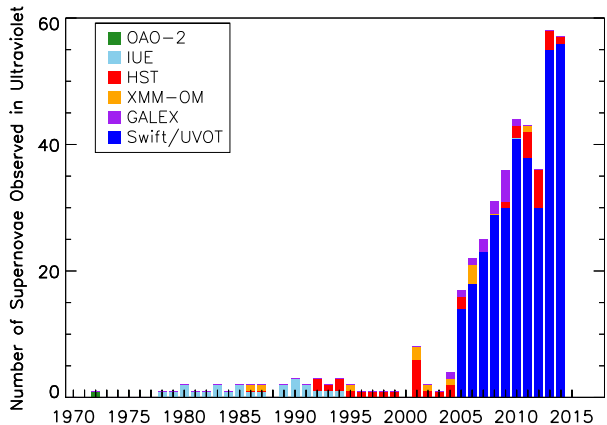} 
\caption[Results]
        {  Supernovae observed in the UV.
 }\label{fig_explosion}    
\end{figure}

The first ultraviolet observation of a supernova  was in 1972 \citep{Holm_etal_1974}.  For three decades, observations with the International Ultraviolet Explorer (IUE), Hubble Space Telescope (HST), the X-ray Multi-Mirror Missions Optical Monitor (XMM-OM), and the Galaxy Evolution Explorer (GALEX) observed only a few supernovae each year (see \citealp{Panagia_2003,Foley_etal_2008_UV,Brown_2009} for a review of these observations and their results).  A dramatic change began in March 2005 when the Swift satellite observed SN~2005am \citep{Brown_etal_2005}.  This started an intensive campaign with Swift's Ultra-Violet Optical Telescope (UVOT; \citealp{Roming_etal_2005}) to characterize the behavior of supernovae in the UV.  The dramatic increase in the number of supernovae observed in the UV is shown in Figure \ref{fig_explosion}.  

\begin{figure*} 
\includegraphics[scale=0.6]{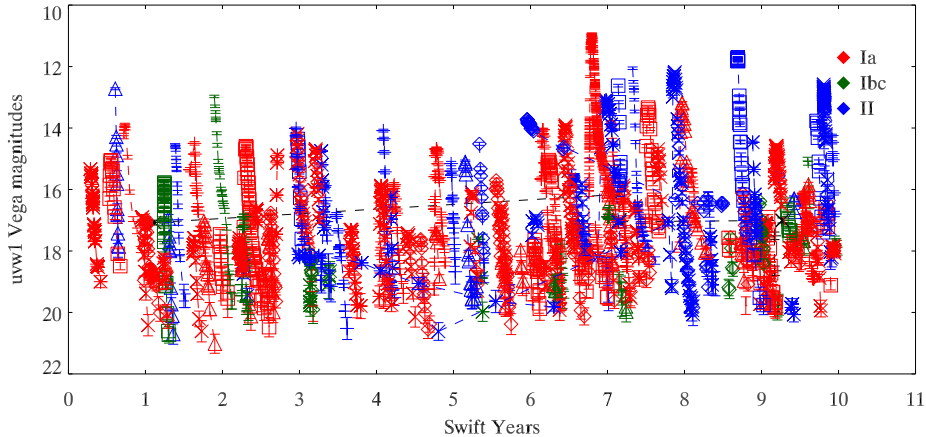} 
\caption[Results]
        {UVOT uvw1 light curves of Swift supernovae over the first ten years.  Dashed lines connect the data points, most noticeable for supernovae detected over a period of several years.  The observed magnitudes show the dynamic range of the Swift supernova observations.
 }\label{fig_lightcurves}    
\end{figure*}

\begin{figure} 
\includegraphics[scale=0.95]{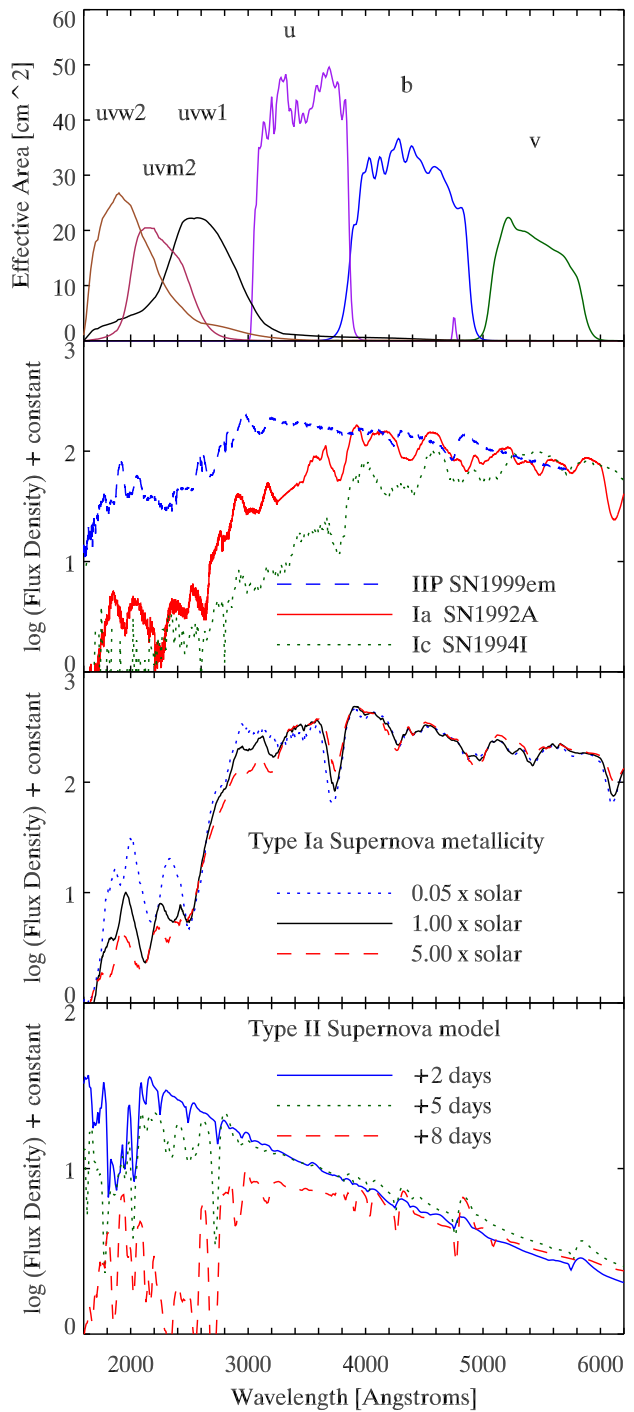} 
\caption[Results]
        {Top Panel: Effective areas of the Swift UVOT filters.
Second Panel: Observed pectra of different types of supernovae showing very different UV-optical spectral shapes.
Third Panel: Theoretical spectra near maximum light of a type Ia supernova showing the different UV behavior resulting from different metallicities (from \citealp{Walker_etal_2012}).
Bottom Panel: Theoretical spectra showing the temporal evolution of a type IIP supernova and the dramatic differences in the UV flux (from \citealp{Dessart_etal_2008}).
 }\label{fig_filters}    
\end{figure}

In its first ten years, Swift has observed over three hundred supernovae.  These cover all major classes and even most of the minor subclasses (some of which were not even identified ten years ago).
Most of those supernovae are not single observations, but the target of a campaign to study the temporal evolution of the UV flux.  A montage of UV light curves is shown in Figure \ref{fig_lightcurves}.  
The UV is very sensitive to many features of the progenitor (e.g. metallicity), explosion (temperature, ejecta density gradients, etc), and environment (reddening, circumstellar interaction, etc.).  Figure \ref{fig_filters} shows the UV region covered by the Swift/UVOT filters compared to the diversity of supernova types, the effect of metallicity on a type Ia supernova spectrum, and the temporal evolution of a type IIP supernova.  The UV is a much more sensitive probe than the optical to many of these effects.
UV light curves for many objects have been published in individual papers along with several compilations \citep{Brown_etal_2009,Milne_etal_2010,Milne_etal_2013,Pritchard_etal_2014}.  The latest release is the Swift Optical/Ultraviolet Supernova Archive (SOUSA; \citealp{Brown_etal_2014_SOUSA}).

Because they are co-pointed, there is a comparable rise in the number of supernovae observed in the X-rays with Swift's X-Ray Telescope (XRT; \citealp{Burrows_etal_2005}.  Though most supernovae are undetected in X-rays, Swift XRT has contributed significantly to the number of supernovae detected at X-rays as well as significant limits on the X-ray production from many others \citep{Koss_Immler_2007,Li_Pun_2011,Ofek_etal_2013,Margutti_etal_2014,Pooley_2014}.

\section{Type Ia Supernovae}

Swift/UVOT observations of type Ia supernovae have dramatically improved the sampling 
of these important, but mysterious explosions, both in the number of events studied and 
in the number of observations per event. The sheer volume of data has been instrumental 
in understanding the basic characteristics of the UV emission as a function of 
the different varieties of type Ia supernovae. Theoretical predictions that the UV wavelength range 
is a vital probe of the explosion physics have been borne out with the recognition 
that the UV emission does not simply trace the optical emission, but instead evolves 
dramatically in UV-optical colors. 
The near-ultraviolet behavior observed with UVOT 
has been found to be similar to the few light curves available from IUE and 
HST, when observing the same or similar 
events \citep{Brown_etal_2005,Kirshner_etal_1993,Wang_etal_2009_05cf}, providing a 
bridge between UVOT photometry and IUE/HST spectroscopy. 
A number of supernovae have also been observed with Swift UVOT's grisms, with spectra quality and observed features comparable to that observed with IUE \citep{Bufano_etal_2009,Foley_etal_2012_09ig,Margutti_etal_2014}.

\citet{Brown_etal_2010} performed the first study of the absolute magnitudes 
of type Ia supernovae in the UV.  They found that the near-UV uvw1 and u filters followed a luminosity-width relation similar to the optical.  The scatter was not significantly larger in the UV than the optical, but both were dominated by observational errors.  As a significant fraction of the errors was due to distances, effort has been made to observe more distant type Ia supernovae with better hubble flow velocities.  
The mid-UV absolute magnitudes showed a large intrinsic scatter.  The nature of that scatter is still under investigation, and there are many more Swift supernovae to be included in future studies.

\subsection{UV-Optical Color Evolution of Groups of type Ia Supernovae} 

The utility of combining UV observations with optical observations has been 
demonstrated by the UV-optical color classes that have become apparent through the 
publication of dozens of type Ia Swift supernovae. Normal type Ia supernovae, the group that 
is important for cosmological measurements, show more diversity in the UV than the optical. The 
UV-optical colors start out very red and become bluest just before the time of 
maximum light in the optical.  After maximum light the colors rapidly becoming redder again. The same 
general behavior is also observed in optical colors, but in the UV the magnitude 
of the color change is greater and the transition from a blue slope to a red slope is far 
more dramatic. \citet{Milne_etal_2013} show a separation of the color curves into 
two groups, having the same temporal shape but offset by color, identified as 
NUV-blue and NUV-red.  These groups are not distinct in the optical, with well 
observed prototypes including the ''golden standard Ia'' 
NUV-red SN~2005cf \citep{Wang_etal_2009_05cf} and the normal but 
NUV-blue SN~2011fe \citep{Brown_etal_2012_11fe,Milne_Brown_2012}. The NUV-blue 
supernovae are also slightly bluer in optical colors, but at a reduced level such that 
they had not been recognized as a separate group before UVOT. 
The search continues to correlate the UV-optical color grouping with features in the 
optical emission. The most promising correlation, to date, has been the presence of 
CII in the early-epoch spectra \citep{Thomas_etal_2011}, a feature attributed to unburned carbon in the 
outermost supernova ejecta. 

The majority ($\sim 2/3$) of nearby normal type Ia supernovae observed by Swift/UVOT 
fall in the NUV-red category \citep{Milne_etal_2015}.  A comparable fraction of 
supernovae observed with near-ultraviolet spectroscopy with 
HST \citep{Maguire_etal_2012} are also NUV-red.  Higher redshift supernovae, where 
the near-ultraviolet light has been redshifted and observed in the optical with 8m-class 
telescopes, have a different distribution with the NUV-blue supernovae 
dominating \citep{Milne_etal_2015}. Indeed, a transition from NUV-red dominance to 
NUV-blue dominance is seen as the samples are separated into redshift bins. 
The color distribution does not shift to the 
blue, but rather, the number of supernovae in the different color regions changes.  
This suggests an evolutionary shift in the relative fractions of the population of 
type Ia progenitors or some other characteristic.

\subsection{Subclasses of Type Ia Supernovae}

The differences described above apply to the optically ``normal'' type Ia supernovae used as standard candles 
to measure cosmology because they are so similar.  
Swift/UVOT has also observed nearly as many non-normal supernovae, and the volume of 
data has allowed the recognition of different emission properties for the subclasses.
\citet{Brown_etal_2014} studied a sample of three super-Chandrasehkar candidates, 
finding that these explosions were UV-bright at all epochs, 
in addition to the previously 
recognized bright optical absolute magnitudes. The additional information from the 
UV emission further strengthens the interpretation that some of these events require 
more $^{56}$Ni then can be produced in a Chandrasekhar-mass explosion. 
That work also presented the UV-optical colors of ``SN 1991T-like" supernovae, often 
interpreted to be a Chandrasekhar-mass explosion that has converted a larger than 
average fraction of the total mass into $^{56}$Ni. Some of the 91T-like supernovae were initially 
quite blue, but transition to normal UV-optical colors with epoch, schematically 
in agreement with the evolution of optical spectra which are peculiar at pre-peak 
epochs, but become fairly normal by optical peak.

\citet{Milne_etal_2010} showed the UV-optical light curves of a number of
optically narrow-peaked supernovae (often referred to as subluminous type Ia supernovae). This group
exhibits a bi-modal distribution, with one group appearing similar to normal
supernovae~Ia in color evolution, while the other is redder at all epochs. The redder
group actually feature UV light curves that are not narrow compared to normal
type Ia supernovae (the optical light curves are narrow and define membership in the
narrow-peaked group). Indeed, \citet{Milne_etal_2010} included the red-group
SN~2005ke in the creation of a mean template from ten type Ia supernovae.

\subsection{Ultraviolet and X-ray Limits on Type Ia Supernova Companions}

Despite their importance as distance indicators, the progenitor systems of type Ia supernovae are not understood (see \citealp{Maoz_etal_2014} for a recent review).  \citet{Kasen_2010} published a testable model perfectly suited to Swift's strengths.  He showed that the interaction of the supernova ejecta should create a shock that would emit strongly in the UV for a few days after the explosion.  

\citet{Brown_etal_2012_shock} started with a sample of twelve type Ia supernovae which had been observed by Swift within ten days of the estimated explosion date.  The numerical models of \citet{Kasen_2010} were found to be an inadequate match to the unshocked supernova emission in the UV, so the analytic models for the shock luminosity itself were used to put limits on the separation distance.  For all of the supernovae, red giant companions were ruled out for most viewing angles.  Factoring in the expected angular distribution, this sample was able to statistically rule out red giants as the companions for more than half of type Ia supernovae.  The extremely nearby SN~2011fe was discovered extremely young and observed by Swift within two days of explosion.  This allowed deep constraints on the brightness of a shock and the separation of a companion \citep{Brown_etal_2012_11fe}. Even a main sequence companion would have to be located at the statistically unlikely bad viewing angle to escape detection according to the \citet{Kasen_2010} models, making the case for a double degenerate progenitor system.  

The UV-bright SN~2011de was adequately fit by the \citet{Kasen_2010} model with a large companion \citep{Brown_2014}.  The uniqueness of the fit, however, is brought into question by the ability of the model to adequately fit the post-maximum decline of several UV-bright Super-Chandrasekhar supernovae \citep{Brown_etal_2014}, some of which had UV spectroscopy or pre-maximum photometry inconsistent with the model predictions.

Swift has a completely independent method of investigating the progenitors  of type Ia supernovae for which the data is nevertheless taken during UV-oriented observations of the supernovae.  Interaction of the supernova ejecta with circumstellar material should create detectable X-rays.  \citet{Russell_Immler_2012} studied the XRT data for a sample of 53 type Ia supernovae, most of which were observed by Swift to study their UV emission.  They computed upper limits to the X-ray luminosity and the corresponding constraint on the mass loss for the individual supernovae and for the whole stacked sample.  For the stacked sample the X-ray limit (0.2-10 keV) is 
L $< 1.7 \times 10^{38}$ erg s$^{-1}$ and the mass loss 
$\dot{M} < 1.1 \times 10^{-6} \,M_\odot yr{-1} × (v w)/(10 km s-1)$.  This rules out red giant stars as the companion stars in a majority of type Ia supernova progenitor systems.

The observations of SN~2011fe turned out to be as deep as the previous samples combined (though the results then only relate to the single system).  \citet{Horesh_etal_2012} measured limits from the first observation of 
 L $< 5 \times 10^{38}$ erg s$^{-1}$ and 
$\dot{M} < 0.2 \times 10^{-6} \,M_\odot yr{-1} × (v w)/(10 km s-1)$.  Combining with radio and Chandra X-ray limits, the mass loss constraint is eventually pushed another two orders of magnitude lower \citep{Margutti_etal_2012}.    
Comparable limits are also placed for the very nearby SN~2014J \citep{Margutti_etal_2014J}.  Taken together, these results suggest a very clean environment around type Ia supernovae.

\section{Core-collapse Supernovae}

Core-collapse supernovae are especially diverse in their UV properties, with hydrogen-dominated type II supernovae typically being UV-bright for significant periods of time and the radioactively-powered type Ib/c supernovae being UV-faint at peak brightness.  \citet{Pritchard_etal_2014} compiled data from fifty of the core-collapse supernovae observed by Swift/UVOT during its first eight years.  The large numbers allow not only the contrasting of different supernova subclasses, but comparisons of the color and luminosity distributions within subclasses.  They also show the large variation the UV makes to the total bolometric luminosity of different supernovae and how it relates to optical observations.

\subsection{Type Ib/c Supernovae and Gamma Ray Burst-Supernova Connection}

Swift is a gamma ray burst mission.  Gamma ray bursts (the $>2$ second long bursts at least) originate from the formation of a black hole during the collapse of a massive star \citep{Thompson_etal_2004,Gehrels_Meszaros_2012}. The rebounding shock also creates an outward expanding supernova explosion (see e.g. \citealp{Woosley_Bloom_2006}).  It would be natural for Swift to study core-collapse supernovae with or without an accompanying gamma ray burst in order to better understand what differences they might have.  
A separate review article in this volume is dedicated to the Gamma-Ray Burst-Supernova connection (Soderberg).

An important event was GRB~060218 and its accompanying supernova 2006aj.  The gamma ray trigger coupled with Swift's fast response allowed the best observations of the shock breakout of a supernova \citep{Campana_etal_2006}.  The duration of the shock breakout is longer than predicted for a compact, massive star, suggestive of an optically thick wind surrounding the supernova progenitor.
The radioactively-powered portion of the supernova is UV-faint.  Once the shock passes through the supernova wind, the supernova cools rapidly and is reheated by radioactive heating.  This second peak is relatively cool and line blanketing from iron-peak elements further suppresses the UV flux.  This prevents the ``supernova bump'' from being observed in the optical at higher redshifts, and when a bump is detected at longer wavelengths its colors are red (see e.g. \citealp{Zeh_etal_2004}).  

While observations of the shock of SN~2006aj was triggered by a gamma ray burst, an even more remarkable shock breakout was observed completely serendipitously.  During observations of the Ib SN~2007uy, an X-ray outburst elsewhere in the same galaxy signaled the explosion of SN~2008D \citep{Soderberg_etal_2008,Modjaz_etal_2009}.  Though heavily reddened, the subsequent UV/optical shock was faintly detected by Swift/UVOT.  

The peculiar supernova 2006jc \citep{Pastorello_etal_2007_06jc,Pastorello_etal_2008_Ibn} inspired a new ``Ibn'' classification (similar to the hydrogen emitting type IIn supernovae to be discussed later) given to supernovae interacting with a helium (rather than hydrogen) dominated circumstellar medium resulting in narrow helium emissions.  It was also UV-bright and featured a brightening in the X-rays when the eject caught up with a shell of ejected material \citep{Immler_etal_2008_06jc}.

\subsection{Type IIb Supernovae}

Type IIb supernovae show hydrogen at early times but the features become weaker and helium appears.  Thus if only a single spectrum is taken, such a supernova could appear to be a typical type II supernova at early times or a type Ib if only observed late.  This transition is caused by an outer layer of hydrogen which contributes at early times but then becomes optically thin.  The effect on the ultraviolet can be understood in terms of the shock breakouts discussed above.  The outer layer of hydrogen will radiate strongly in the UV but fade as it cools.  The fading is slower than for the smaller stripped envelope core-collapse supernovae, so the shock is visible for a longer time.  
Early discoveries and prompt Swift/UVOT observations are thus able to see fading UV emission from this adiabatic cooling.   These shock breakouts have a range of brightness and durations corresponding to the size of the hydrogen envelope, ranging from faint (SN~2008ax; \citealp{Roming_etal_2009_08ax}) to brighter (SN~2011dh; \citealp{Arcavi_etal_2011,Marion_etal_2014,Ergon_etal_2014}) and quite bright (SN~2013df; \citealp{Morales_etal_2014}).  

\subsection{Type IIP Supernovae}

Type IIP supernovae have large hydrogen envelopes, so their UV flux would be expected to last longer than the stripped-envelope core-collapse supernovae discussed above.  Swift/UVOT observations of SN~2005cs confirmed an early, bright UV luminosity and were the first to show the temporal evolution of a IIP \citep{Brown_etal_2007_05cs}.  These observations showed a monotonic flux decrease in all the UV filters while the V-band magnitude remained nearly constant (the defining feature of a type II-``P''lateau).  The first epoch showed a strong UV-continuum, consistent with a temperature of 15,750 K and line absorption from NiIII and FeIII.  Radiative transfer modeling on SNe~2005cs and 2006bp showed the UV drop is driven by the falling temperature, with ionization shifts (e.g. NiIII to NiII and FeIII to FeII) and strengthening line absorption also contributing \citep{Dessart_etal_2008}.   Once the temperature stabilizes, the UV light curves flatten out, an effect best seen in the late observations of SN~2012aw \citep{Bayless_etal_2013}.  

\citet{Pejcha_Prieto_2015} incorporated Swift/UVOT observations of type IIP supernovae into an empirical model of the photospheric radius and temperature variations
Back on the theoretical side, \citet{Bayless_etal_2014} modeled the effect the mass of the hydrogen envelope has on the shock breakout and subsequent light curve evolution.  Removing hydrogen results in a quicker drop off in the UV and optical flux, broadly following the different classes described above.

\subsection{Type IIn Supernovae}

Type IIn supernovae are identified by narrow hydrogen emission lines, rising from circumstellar material which has been shocked by the supernova ejecta.  They show a wide variety of UV behavior.   SN~2007pk began UV bright but reddened quickly, similar to type IIP supernovae \citep{Pritchard_etal_2012}.  SN~2010jl was extremely UV-bright and slowly faded in the UV and optical with a nearly constant color, while in the X-rays it brightened after two hundred days and has remained bright \citep{Ofek_etal_2014}.  SN~2011ht began red, became bluer as it brightened over 30 days, and then reddened as it faded \citep{Roming_etal_2012}.  SN~2009ip has a similar, though much more rapid, evolution \citep{Margutti_etal_2014}.  It was unique in that it was first detected as an outburst of some sort several years before a much more luminous outburst.  Whether the later outbursts of SNe 2009ip and 2011ht were terminal, however, is still under dispute and observation \citep{Humphreys_etal_2012,Pastorello_etal_2013,Fraser_etal_2013}.  The characteristic emission of a type IIn supernova comes from an optically thick shell, resulting in a form of stellar amnesia where the external observables are the same regardless of the underlying explosion.  Thus the collision of shells from outbursts of a luminous blue variable can result in a very luminous event even if the star has not terminally exploded.   This emission can also come at different times with a transient being first identified as a IIn and later appearing like a normal IIP (PTF11iqb; \citealp{Smith_etal_2015}) or objects first classified as IIb which later interacted with their surrounding medium, spectroscopically resembling type IIn supernovae and emitting UV radiation for years (SNe 2005ip and 2006jd; \citealp{Stritzinger_etal_2012}).  The bright UV emission of type IIn supernovae allows them to be detected at very large distances, with the highest at a redshift of $\sim$2 \citep{Cooke_etal_2009}.  Swift UV observations of nearby type IIn supernovae will provide an important comparison set for these distant objects.

\subsection{Superluminous Supernovae}

A new class of Superluminous Supernovae (SLSNe) has only recently been discovered which are up to 50 times more luminous than the  Type Ia supernovae used for cosmological measurements \citep{Pastorello_etal_2010, Quimby_etal_2011, Gal-Yam_2012}.   Their observed behavior is varied, though, and the $\sim$20 well studied examples have been divided into separate groups (see \citealp{Gal-Yam_2012} for a review).  Analogous to classical supernova typing, SLSNe I have no hydrogen in their spectra, and they may be related to Ic supernovae \citep{Pastorello_etal_2010}.  Their high peak luminosity could be from a large mass of radioactive Ni; the light curves, however, are too narrow for the required mass.  \citet{Inserra_etal_2013} modelled the light curves of five SLSNe I with energy coming from the spin-down of a magnetic neutron star.  \citet{Gal-Yam_2012} suggested a SLSN-R class which also shows no hydrogen but are powered by radioactive decay.  Their progenitors must be extremely massive, with their explosion possibly triggered by pair-instability as was argued for SN~2007bi \citep{Gal-Yam_etal_2009}, though similar objects are inconsistent with such a model \citep{Nicholl_etal_2013}.  The overlap between these two groups is still uncertain, especially given the small numbers of objects.

In addition to their higher optical luminosity, an important difference between SLSNe and type Ia supernovae is their UV luminosity. For reasons not yet understood, SLSNe I do not show the strong metal line blanketing which suppresses the UV flux in other type I supernovae.  Type II SLSNe show a very strong rise in flux to shorter wavelengths, similar to the hot photospheres of the classical hydrogen-dominated type II supernovae.  For the earliest observations of SN~2008es the wavelength of the peak flux was shortward of the Swift UV observations (rest wavelength of $\sim$1500 \AA; \citealp{Gezari_etal_2009}.  Thus UV observations are important for measuring the total luminosity and constraining the temperature.  It is also key to understanding the explosion mechanism(s).  From an observational standpoint, the high UV flux makes it much easier to detect these supernovae at higher redshifts because the spectral shape (and its shift with redshift into the observed bands) helps rather than hurts. The current most distant SLSN was discovered at z=3.9 \citep{Cooke_etal_2012}.  

\begin{figure} 
\includegraphics[scale=0.95]{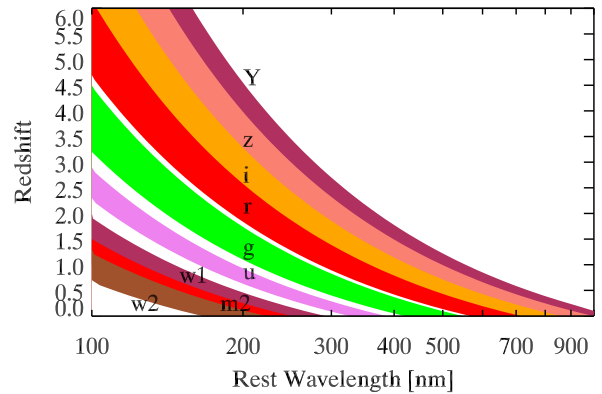} 
\caption[Results]
        {Redshift versus rest-frame wavelengths covered by different filters.  The redshifting of light from distant supernovae requires knowledge of shorter rest-frame wavelengths of nearby supernovae in order to compare how their properties might change over the history of the universe.
 }\label{fig_redshift}    
\end{figure}

While the high luminosity makes them detectable at higher redshifts, the redshifting of light makes it harder to systematically compare SLSNe at the same rest-frame wavelengths.    
Most observations are made in the gri bands which cover exclusively space-UV rest-wavelengths beyond a redshift of 1.7.  For the SLSN at z=3.9 the rest frame gri corresponds to rest wavelengths of 1000-3000 \AA.  
As shown in Figure \ref{fig_redshift}, observer-frame optical observations trace out tracks of wavelengths that move to shorter wavelengths as the redshift increases.  At each redshift a different rest-frame wavelength range is probed by optical observations.  Observer-frame UV observations probe parallel tracks of rest-frame wavelengths, allowing a comparison with higher redshift optical observations.  These are needed to compare SLSNe at the rest wavelengths from which the bulk of their emission is emitted.

\section{The Future of Swift Supernovae}

The large diversity of Swift supernova observations have raised questions that were not even being asked when Swift launched.  While the quantity of Swift supernova observations is incredible, we hope to continue many more observations in the future.  Larger numbers of type Ia supernovae, for example, will allow statistically significant numbers of events to be studied with different sample cuts and regressions.  This will aid in breaking the degeneracies of metallicity, reddening, density gradients, explosion differences which all affect the UV flux.  

Future samples of supernovae will also push to younger times where the rapid response of Swift is even more important and can open up new discover channels.  The upcoming Zwicky Transient Facility and other near-term projects focusing on rapid cadence supernova searches will provide excellent targets for Swift follow up.  Swift's efficient and responsive Target-of-Opportunity request system allows the supernova community to request observation, while the prompt and public release of data allows for timely data analysis by anyone.

The dissemination of the Swift data allows for more comparisons with complementary data sets, such as spectral velocity and abundance measurements, host galaxy properties, and other multi-wavelength parameters.  The Swift/UVOT observations will be the local benchmark of rest-frame UV characteristics against which to compare the high redshift supernovae which are observed by past, present, and future optical telescopes.

For years the theoretical understanding of the UV flux of supernovae has been data starved.  With the exception of SN~1987A, most UV observations were limited to a few epochs near maximum light.  Rather than matching a single object at a particular epoch, theorists can now test whether their model can be appropriately varied to match the distribution of objects and how their differences change with time.  For type Ia supernovae, constraints on the explosions and environments may improve their utility as standard candles.  For core-collapse supernovae, the Swift data will teach us about mass loss during the final stages of life and death of massive stars.  Examples of many rare subclasses now have UV and X-ray data to constrain the nature of these explosions.  This deeper understanding of the wide diversity of stellar explosions will be one of Swift's last legacies.

\section{Acknowledgements}

P.J.B. is supported by the Mitchell Postdoctoral Fellowship, NASA ADAP grant NNX13AF35G, and NASA Swift GI grant NNX14AC52G.



\begin{thebibliography}{}
\expandafter\ifx\csname natexlab\endcsname\relax\def\natexlab#1{#1}\fi

\bibitem[{{Arcavi} {et~al.}(2011){Arcavi}, {Gal-Yam}, {Yaron}, {Sternberg},
  {Rabinak}, {Waxman}, {Kasliwal}, {Quimby}, {Ofek}, {Horesh}, {Kulkarni},
  {Filippenko}, {Silverman}, {Cenko}, {Li}, {Bloom}, {Sullivan}, {Nugent},
  {Poznanski}, {Gorbikov}, {Fulton}, {Howell}, {Bersier}, {Riou},
  {Lamotte-Bailey}, {Griga}, {Cohen}, {Hachinger}, {Polishook}, {Xu},
  {Ben-Ami}, {Manulis}, {Walker}, {Maguire}, {Pan}, {Matheson}, {Mazzali},
  {Pian}, {Fox}, {Gehrels}, {Law}, {James}, {Marchant}, {Smith}, {Mottram},
  {Barnsley}, {Kandrashoff}, \& {Clubb}}]{Arcavi_etal_2011}
{Arcavi}, I., {Gal-Yam}, A., {Yaron}, O., {et~al.} 2011, ApJL, 742, L18

\bibitem[{{Bayless} {et~al.}(2014){Bayless}, {Even}, {Frey}, {Fryer}, {Roming},
  \& {Young}}]{Bayless_etal_2014}
{Bayless}, A.~J., {Even}, W., {Frey}, L.~H., {et~al.} 2014, 
  arXiv:1401.4449

\bibitem[{{Bayless} {et~al.}(2013){Bayless}, {Pritchard}, {Roming}, {Kuin},
  {Brown}, {Botticella}, {Dall'Ora}, {Frey}, {Even}, {Fryer}, {Maund}, \&
  {Fraser}}]{Bayless_etal_2013}
{Bayless}, A.~J., {Pritchard}, T.~A., {Roming}, P.~W.~A., {et~al.} 2013, ApJL,
  764, L13

\bibitem[{{Brown}(2009)}]{Brown_2009}
{Brown}, P.~J. 2009, PhD thesis, The Pennsylvania State University

\bibitem[{{Brown}(2014)}]{Brown_2014}
---. 2014, ApJL, 796, L18

\bibitem[{{Brown} {et~al.}(2014{\natexlab{a}}){Brown}, {Breeveld}, {Holland},
  {Kuin}, \& {Pritchard}}]{Brown_etal_2014_SOUSA}
{Brown}, P.~J., {Breeveld}, A.~A., {Holland}, S., {Kuin}, P., \& {Pritchard},
  T. 2014{\natexlab{a}}, A\&SS, 354, 89

\bibitem[{{Brown} {et~al.}(2012{\natexlab{a}}){Brown}, {Dawson}, {Harris},
  {Olmstead}, {Milne}, \& {Roming}}]{Brown_etal_2012_shock}
{Brown}, P.~J., {Dawson}, K.~S., {Harris}, D.~W., {et~al.} 2012{\natexlab{a}},
  ApJ, 749, 18

\bibitem[{{Brown} {et~al.}(2005){Brown}, {Holland}, {James}, {Milne}, {Roming},
  {Mason}, {Page}, {Beardmore}, {Burrows}, {Morgan}, {Gronwall}, {Blustin},
  {Boyd}, {Still}, {Breeveld}, {de Pasquale}, {Hunsberger}, {Ivanushkina},
  {Landsman}, {McGowan}, {Poole}, {Rosen}, {Schady}, \&
  {Gehrels}}]{Brown_etal_2005}
{Brown}, P.~J., {Holland}, S.~T., {James}, C., {et~al.} 2005, ApJ, 635, 1192

\bibitem[{{Brown} {et~al.}(2007){Brown}, {Dessart}, {Holland}, {Immler},
  {Landsman}, {Blondin}, {Blustin}, {Breeveld}, {Dewangan}, {Gehrels},
  {Hutchins}, {Kirshner}, {Mason}, {Mazzali}, {Milne}, {Modjaz}, \&
  {Roming}}]{Brown_etal_2007_05cs}
{Brown}, P.~J., {Dessart}, L., {Holland}, S.~T., {et~al.} 2007, ApJ, 659, 1488

\bibitem[{{Brown} {et~al.}(2009){Brown}, {Holland}, {Immler}, {Milne},
  {Roming}, {Gehrels}, {Nousek}, {Panagia}, {Still}, \& {Vanden
  Berk}}]{Brown_etal_2009}
{Brown}, P.~J., {Holland}, S.~T., {Immler}, S., {et~al.} 2009, AJ, 137, 4517

\bibitem[{{Brown} {et~al.}(2010){Brown}, {Roming}, {Milne}, {Bufano},
  {Ciardullo}, {Elias-Rosa}, {Filippenko}, {Foley}, {Gehrels}, {Gronwall},
  {Hicken}, {Holland}, {Hoversten}, {Immler}, {Kirshner}, {Li}, {Mazzali},
  {Phillips}, {Pritchard}, {Still}, {Turatto}, \& {Vanden
  Berk}}]{Brown_etal_2010}
{Brown}, P.~J., {Roming}, P.~W.~A., {Milne}, P., {et~al.} 2010, ApJ, 721, 1608

\bibitem[{{Brown} {et~al.}(2012{\natexlab{b}}){Brown}, {Dawson}, {de Pasquale},
  {Gronwall}, {Holland}, {Immler}, {Kuin}, {Mazzali}, {Milne}, {Oates}, \&
  {Siegel}}]{Brown_etal_2012_11fe}
{Brown}, P.~J., {Dawson}, K.~S., {de Pasquale}, M., {et~al.}
  2012{\natexlab{b}}, ApJ, 753, 22

\bibitem[{{Brown} {et~al.}(2014{\natexlab{b}}){Brown}, {Kuin}, {Scalzo},
  {Smitka}, {de Pasquale}, {Holland}, {Krisciunas}, {Milne}, \&
  {Wang}}]{Brown_etal_2014}
{Brown}, P.~J., {Kuin}, P., {Scalzo}, R., {et~al.} 2014{\natexlab{b}}, ApJ,
  787, 29

\bibitem[{{Bufano} {et~al.}(2009){Bufano}, {Immler}, {Turatto}, {Landsman},
  {Brown}, {Benetti}, {Cappellaro}, {Holland}, {Mazzali}, {Milne}, {Panagia},
  {Pian}, {Roming}, {Zampieri}, {Breeveld}, \& {Gehrels}}]{Bufano_etal_2009}
{Bufano}, F., {Immler}, S., {Turatto}, M., {et~al.} 2009, ApJ, 700, 1456

\bibitem[{{Burrows} {et~al.}(2005){Burrows}, {Hill}, {Nousek}, {Kennea},
  {Wells}, {Osborne}, {Abbey}, {Beardmore}, {Mukerjee}, {Short}, {Chincarini},
  {Campana}, {Citterio}, {Moretti}, {Pagani}, {Tagliaferri}, {Giommi},
  {Capalbi}, {Tamburelli}, {Angelini}, {Cusumano}, {Br{\"a}uninger}, {Burkert},
  \& {Hartner}}]{Burrows_etal_2005}
{Burrows}, D.~N., {Hill}, J.~E., {Nousek}, J.~A., {et~al.} 2005, Space Science
  Reviews, 120, 165

\bibitem[{{Campana} {et~al.}(2006){Campana}, {Mangano}, {Blustin}, {Brown},
  {Burrows}, {Chincarini}, {Cummings}, {Cusumano}, {Della Valle}, {Malesani},
  {M{\'e}sz{\'a}ros}, {Nousek}, {Page}, {Sakamoto}, {Waxman}, {Zhang}, {Dai},
  {Gehrels}, {Immler}, {Marshall}, {Mason}, {Moretti}, {O'Brien}, {Osborne},
  {Page}, {Romano}, {Roming}, {Tagliaferri}, {Cominsky}, {Giommi}, {Godet},
  {Kennea}, {Krimm}, {Angelini}, {Barthelmy}, {Boyd}, {Palmer}, {Wells}, \&
  {White}}]{Campana_etal_2006}
{Campana}, S., {Mangano}, V., {Blustin}, A.~J., {et~al.} 2006, Nature, 442,
  1008

\bibitem[{{Cooke} {et~al.}(2009){Cooke}, {Sullivan}, {Barton}, {Bullock},
  {Carlberg}, {Gal-Yam}, \& {Tollerud}}]{Cooke_etal_2009}
{Cooke}, J., {Sullivan}, M., {Barton}, E.~J., {et~al.} 2009, Nature, 460, 237

\bibitem[{{Cooke} {et~al.}(2012){Cooke}, {Sullivan}, {Gal-Yam}, {Barton},
  {Carlberg}, {Ryan-Weber}, {Horst}, {Omori}, \&
  {D{\'{\i}}az}}]{Cooke_etal_2012}
{Cooke}, J., {Sullivan}, M., {Gal-Yam}, A., {et~al.} 2012, Nature, 491, 228

\bibitem[{{Dessart} {et~al.}(2008){Dessart}, {Blondin}, {Brown}, {Hicken},
  {Hillier}, {Holland}, {Immler}, {Kirshner}, {Milne}, {Modjaz}, \&
  {Roming}}]{Dessart_etal_2008}
{Dessart}, L., {Blondin}, S., {Brown}, P.~J., {et~al.} 2008, ApJ, 675, 644

\bibitem[{{Ergon} {et~al.}(2014){Ergon}, {Sollerman}, {Fraser}, {Pastorello},
  {Taubenberger}, {Elias-Rosa}, {Bersten}, {Jerkstrand}, {Benetti},
  {Botticella}, {Fransson}, {Harutyunyan}, {Kotak}, {Smartt}, {Valenti},
  {Bufano}, {Cappellaro}, {Fiaschi}, {Howell}, {Kankare}, {Magill}, {Mattila},
  {Maund}, {Naves}, {Ochner}, {Ruiz}, {Smith}, {Tomasella}, \&
  {Turatto}}]{Ergon_etal_2014}
{Ergon}, M., {Sollerman}, J., {Fraser}, M., {et~al.} 2014, AAP, 562, A17

\bibitem[{{Foley} {et~al.}(2008){Foley}, {Filippenko}, \&
  {Jha}}]{Foley_etal_2008_UV}
{Foley}, R.~J., {Filippenko}, A.~V., \& {Jha}, S.~W. 2008, ApJ, 686, 117

\bibitem[{{Foley} {et~al.}(2012){Foley}, {Challis}, {Filippenko},
  {Ganeshalingam}, {Landsman}, {Li}, {Marion}, {Silverman}, {Beaton},
  {Bennert}, {Cenko}, {Childress}, {Guhathakurta}, {Jiang}, {Kalirai},
  {Kirshner}, {Stockton}, {Tollerud}, {Vink{\'o}}, {Wheeler}, \&
  {Woo}}]{Foley_etal_2012_09ig}
{Foley}, R.~J., {Challis}, P.~J., {Filippenko}, A.~V., {et~al.} 2012, ApJ, 744,
  38

\bibitem[{{Fraser} {et~al.}(2013){Fraser}, {Inserra}, {Jerkstrand}, {Kotak},
  {Pignata}, {Benetti}, {Botticella}, {Bufano}, {Childress}, {Mattila},
  {Pastorello}, {Smartt}, {Turatto}, {Yuan}, {Anderson}, {Bayliss}, {Bauer},
  {Chen}, {F{\"o}rster Bur{\'o}n}, {Gal-Yam}, {Haislip}, {Knapic}, {Le
  Guillou}, {Marchi}, {Mazzali}, {Molinaro}, {Moore}, {Reichart}, {Smareglia},
  {Smith}, {Sternberg}, {Sullivan}, {Tak{\'a}ts}, {Tucker}, {Valenti}, {Yaron},
  {Young}, \& {Zhou}}]{Fraser_etal_2013}
{Fraser}, M., {Inserra}, C., {Jerkstrand}, A., {et~al.} 2013, MNRAS, 433, 1312

\bibitem[{{Gal-Yam}(2012)}]{Gal-Yam_2012}
{Gal-Yam}, A. 2012, Science, 337, 927

\bibitem[{{Gal-Yam} {et~al.}(2009){Gal-Yam}, {Mazzali}, {Ofek}, {Nugent},
  {Kulkarni}, {Kasliwal}, {Quimby}, {Filippenko}, {Cenko}, {Chornock},
  {Waldman}, {Kasen}, {Sullivan}, {Beshore}, {Drake}, {Thomas}, {Bloom},
  {Poznanski}, {Miller}, {Foley}, {Silverman}, {Arcavi}, {Ellis}, \&
  {Deng}}]{Gal-Yam_etal_2009}
{Gal-Yam}, A., {Mazzali}, P., {Ofek}, E.~O., {et~al.} 2009, Nature, 462, 624

\bibitem[{{Gehrels} \& {M{\'e}sz{\'a}ros}(2012)}]{Gehrels_Meszaros_2012}
{Gehrels}, N., \& {M{\'e}sz{\'a}ros}, P. 2012, Science, 337, 932

\bibitem[{{Gehrels} {et~al.}(2004){Gehrels}, {Chincarini}, {Giommi}, {Mason},
  {Nousek}, {Wells}, {White}, {Barthelmy}, {Burrows}, {Cominsky}, {Hurley},
  {Marshall}, {M{\'e}sz{\'a}ros}, {Roming}, {Angelini}, {Barbier}, {Belloni},
  {Campana}, {Caraveo}, {Chester}, {Citterio}, {Cline}, {Cropper}, {Cummings},
  {Dean}, {Feigelson}, {Fenimore}, {Frail}, {Fruchter}, {Garmire}, {Gendreau},
  {Ghisellini}, {Greiner}, {Hill}, {Hunsberger}, {Krimm}, {Kulkarni}, {Kumar},
  {Lebrun}, {Lloyd-Ronning}, {Markwardt}, {Mattson}, {Mushotzky}, {Norris},
  {Osborne}, {Paczynski}, {Palmer}, {Park}, {Parsons}, {Paul}, {Rees},
  {Reynolds}, {Rhoads}, {Sasseen}, {Schaefer}, {Short}, {Smale}, {Smith},
  {Stella}, {Tagliaferri}, {Takahashi}, {Tashiro}, {Townsley}, {Tueller},
  {Turner}, {Vietri}, {Voges}, {Ward}, {Willingale}, {Zerbi}, \&
  {Zhang}}]{Gehrels_etal_2004}
{Gehrels}, N., {Chincarini}, G., {Giommi}, P., {et~al.} 2004, ApJ, 611, 1005

\bibitem[{{Gezari} {et~al.}(2009){Gezari}, {Halpern}, {Grupe}, {Yuan},
  {Quimby}, {McKay}, {Chamarro}, {Sisson}, {Akerlof}, {Wheeler}, {Brown},
  {Cenko}, {Rau}, {Djordjevic}, \& {Terndrup}}]{Gezari_etal_2009}
{Gezari}, S., {Halpern}, J.~P., {Grupe}, D., {et~al.} 2009, ApJ, 690, 1313

\bibitem[{{Holm} {et~al.}(1974){Holm}, {Wu}, \& {Caldwell}}]{Holm_etal_1974}
{Holm}, A.~V., {Wu}, C.-C., \& {Caldwell}, J.~J. 1974, PASP, 86, 296

\bibitem[{{Horesh} {et~al.}(2012){Horesh}, {Kulkarni}, {Fox}, {Carpenter},
  {Kasliwal}, {Ofek}, {Quimby}, {Gal-Yam}, {Cenko}, {de Bruyn}, {Kamble},
  {Wijers}, {van der Horst}, {Kouveliotou}, {Podsiadlowski}, {Sullivan},
  {Maguire}, {Howell}, {Nugent}, {Gehrels}, {Law}, {Poznanski}, \&
  {Shara}}]{Horesh_etal_2012}
{Horesh}, A., {Kulkarni}, S.~R., {Fox}, D.~B., {et~al.} 2012, ApJ, 746, 21

\bibitem[{{Humphreys} {et~al.}(2012){Humphreys}, {Davidson}, {Jones}, {Pogge},
  {Grammer}, {Prieto}, \& {Pritchard}}]{Humphreys_etal_2012}
{Humphreys}, R.~M., {Davidson}, K., {Jones}, T.~J., {et~al.} 2012, ApJ, 760, 93

\bibitem[{{Immler} {et~al.}(2008){Immler}, {Modjaz}, {Landsman}, {Bufano},
  {Brown}, {Milne}, {Dessart}, {Holland}, {Koss}, {Pooley}, {Kirshner},
  {Filippenko}, {Panagia}, {Chevalier}, {Mazzali}, {Gehrels}, {Petre},
  {Burrows}, {Nousek}, {Roming}, {Pian}, {Soderberg}, \&
  {Greiner}}]{Immler_etal_2008_06jc}
{Immler}, S., {Modjaz}, M., {Landsman}, W., {et~al.} 2008, ApJL, 674, L85

\bibitem[{{Inserra} {et~al.}(2013){Inserra}, {Smartt}, {Jerkstrand}, {Valenti},
  {Fraser}, {Wright}, {Smith}, {Chen}, {Kotak}, {Pastorello}, {Nicholl},
  {Bresolin}, {Kudritzki}, {Benetti}, {Botticella}, {Burgett}, {Chambers},
  {Ergon}, {Flewelling}, {Fynbo}, {Geier}, {Hodapp}, {Howell}, {Huber},
  {Kaiser}, {Leloudas}, {Magill}, {Magnier}, {McCrum}, {Metcalfe}, {Price},
  {Rest}, {Sollerman}, {Sweeney}, {Taddia}, {Taubenberger}, {Tonry},
  {Wainscoat}, {Waters}, \& {Young}}]{Inserra_etal_2013}
{Inserra}, C., {Smartt}, S.~J., {Jerkstrand}, A., {et~al.} 2013, ApJ, 770, 128

\bibitem[{{Kasen}(2010)}]{Kasen_2010}
{Kasen}, D. 2010, ApJ, 708, 1025

\bibitem[{{Kirshner} {et~al.}(1993){Kirshner}, {Jeffery}, {Leibundgut},
  {Challis}, {Sonneborn}, {Phillips}, {Suntzeff}, {Smith}, {Winkler}, {Winge},
  {Hamuy}, {Hunter}, {Roth}, {Blades}, {Branch}, {Chevalier}, {Fransson},
  {Panagia}, {Wagoner}, {Wheeler}, \& {Harkness}}]{Kirshner_etal_1993}
{Kirshner}, R.~P., {Jeffery}, D.~J., {Leibundgut}, B., {et~al.} 1993, ApJ, 415,
  589

\bibitem[{{Koss} \& {Immler}(2007)}]{Koss_Immler_2007}
{Koss}, M., \& {Immler}, S. 2007, in American Institute of Physics Conference
  Series, Vol. 937, Supernova 1987A: 20 Years After: Supernovae and Gamma-Ray
  Bursters, ed. S.~{Immler}, K.~{Weiler}, \& R.~{McCray}, 436--439

\bibitem[{{Li} \& {Pun}(2011)}]{Li_Pun_2011}
{Li}, K.~L., \& {Pun}, C.~S.~J. 2011, arXiv:1109.0981

\bibitem[{{Maguire} {et~al.}(2012){Maguire}, {Sullivan}, {Ellis}, {Nugent},
  {Howell}, {Gal-Yam}, {Cooke}, {Mazzali}, {Pan}, {Dilday}, {Thomas}, {Arcavi},
  {Ben-Ami}, {Bersier}, {Bianco}, {Fulton}, {Hook}, {Horesh}, {Hsiao}, {James},
  {Podsiadlowski}, {Walker}, {Yaron}, {Kasliwal}, {Laher}, {Law}, {Ofek},
  {Poznanski}, \& {Surace}}]{Maguire_etal_2012}
{Maguire}, K., {Sullivan}, M., {Ellis}, R.~S., {et~al.} 2012, MNRAS, 426, 2359

\bibitem[{{Maoz} {et~al.}(2014){Maoz}, {Mannucci}, \&
  {Nelemans}}]{Maoz_etal_2014}
{Maoz}, D., {Mannucci}, F., \& {Nelemans}, G. 2014, ARAA, 52, 107

\bibitem[{{Margutti} {et~al.}(2014{\natexlab{a}}){Margutti}, {Parrent},
  {Kamble}, {Soderberg}, {Foley}, {Milisavljevic}, {Drout}, \&
  {Kirshner}}]{Margutti_etal_2014J}
{Margutti}, R., {Parrent}, J., {Kamble}, A., {et~al.} 2014{\natexlab{a}}, ApJ,
  790, 52

\bibitem[{{Margutti} {et~al.}(2012){Margutti}, {Soderberg}, {Chomiuk},
  {Chevalier}, {Hurley}, {Milisavljevic}, {Foley}, {Hughes}, {Slane},
  {Fransson}, {Moe}, {Barthelmy}, {Boynton}, {Briggs}, {Connaughton}, {Costa},
  {Cummings}, {Del Monte}, {Enos}, {Fellows}, {Feroci}, {Fukazawa}, {Gehrels},
  {Goldsten}, {Golovin}, {Hanabata}, {Harshman}, {Krimm}, {Litvak},
  {Makishima}, {Marisaldi}, {Mitrofanov}, {Murakami}, {Ohno}, {Palmer},
  {Sanin}, {Starr}, {Svinkin}, {Takahashi}, {Tashiro}, {Terada}, \&
  {Yamaoka}}]{Margutti_etal_2012}
{Margutti}, R., {Soderberg}, A.~M., {Chomiuk}, L., {et~al.} 2012, ApJ, 751, 134

\bibitem[{{Margutti} {et~al.}(2014{\natexlab{b}}){Margutti}, {Milisavljevic},
  {Soderberg}, {Chornock}, {Zauderer}, {Murase}, {Guidorzi}, {Sanders}, {Kuin},
  {Fransson}, {Levesque}, {Chandra}, {Berger}, {Bianco}, {Brown}, {Challis},
  {Chatzopoulos}, {Cheung}, {Choi}, {Chomiuk}, {Chugai}, {Contreras}, {Drout},
  {Fesen}, {Foley}, {Fong}, {Friedman}, {Gall}, {Gehrels}, {Hjorth}, {Hsiao},
  {Kirshner}, {Im}, {Leloudas}, {Lunnan}, {Marion}, {Martin}, {Morrell},
  {Neugent}, {Omodei}, {Phillips}, {Rest}, {Silverman}, {Strader},
  {Stritzinger}, {Szalai}, {Utterback}, {Vinko}, {Wheeler}, {Arnett},
  {Campana}, {Chevalier}, {Ginsburg}, {Kamble}, {Roming}, {Pritchard}, \&
  {Stringfellow}}]{Margutti_etal_2014}
{Margutti}, R., {Milisavljevic}, D., {Soderberg}, A.~M., {et~al.}
  2014{\natexlab{b}}, ApJ, 780, 21

\bibitem[{{Marion} {et~al.}(2014){Marion}, {Vinko}, {Kirshner}, {Foley},
  {Berlind}, {Bieryla}, {Bloom}, {Calkins}, {Challis}, {Chevalier}, {Chornock},
  {Culliton}, {Curtis}, {Esquerdo}, {Everett}, {Falco}, {France}, {Fransson},
  {Friedman}, {Garnavich}, {Leibundgut}, {Meyer}, {Smith}, {Soderberg},
  {Sollerman}, {Starr}, {Szklenar}, {Takats}, \& {Wheeler}}]{Marion_etal_2014}
{Marion}, G.~H., {Vinko}, J., {Kirshner}, R.~P., {et~al.} 2014, ApJ, 781, 69

\bibitem[{{Milne} \& {Brown}(2012)}]{Milne_Brown_2012}
{Milne}, P.~A., \& {Brown}, P.~J. 2012, in The Extreme and Variable High Energy
  Sky, Proceedings of Science, 1--5

\bibitem[{{Milne} {et~al.}(2013){Milne}, {Brown}, {Roming}, {Bufano}, \&
  {Gehrels}}]{Milne_etal_2013}
{Milne}, P.~A., {Brown}, P.~J., {Roming}, P.~W.~A., {Bufano}, F., \& {Gehrels},
  N. 2013, ApJ, 779, 23

\bibitem[{{Milne} {et~al.}(2015){Milne}, {Foley}, {Brown}, \&
  {Narayan}}]{Milne_etal_2015}
{Milne}, P.~A., {Foley}, R.~J., {Brown}, P.~J., \& {Narayan}, G. 2015, ApJ,
  803, 20

\bibitem[{{Milne} {et~al.}(2010){Milne}, {Brown}, {Roming}, {Holland},
  {Immler}, {Filippenko}, {Ganeshalingam}, {Li}, {Stritzinger}, {Phillips},
  {Hicken}, {Kirshner}, {Challis}, {Mazzali}, {Schmidt}, {Bufano}, {Gehrels},
  \& {Vanden Berk}}]{Milne_etal_2010}
{Milne}, P.~A., {Brown}, P.~J., {Roming}, P.~W.~A., {et~al.} 2010, ApJ, 721,
  1627

\bibitem[{{Modjaz} {et~al.}(2009){Modjaz}, {Li}, {Butler}, {Chornock},
  {Perley}, {Blondin}, {Bloom}, {Filippenko}, {Kirshner}, {Kocevski},
  {Poznanski}, {Hicken}, {Foley}, {Stringfellow}, {Berlind}, {Barrado y
  Navascues}, {Blake}, {Bouy}, {Brown}, {Challis}, {Chen}, {de Vries},
  {Dufour}, {Falco}, {Friedman}, {Ganeshalingam}, {Garnavich}, {Holden},
  {Illingworth}, {Lee}, {Liebert}, {Marion}, {Olivier}, {Prochaska},
  {Silverman}, {Smith}, {Starr}, {Steele}, {Stockton}, {Williams}, \&
  {Wood-Vasey}}]{Modjaz_etal_2009}
{Modjaz}, M., {Li}, W., {Butler}, N., {et~al.} 2009, ApJ, 702, 226

\bibitem[{{Morales-Garoffolo} {et~al.}(2014){Morales-Garoffolo}, {Elias-Rosa},
  {Benetti}, {Taubenberger}, {Cappellaro}, {Pastorello}, {Klauser}, {Valenti},
  {Howerton}, {Ochner}, {Schramm}, {Siviero}, {Tartaglia}, \&
  {Tomasella}}]{Morales_etal_2014}
{Morales-Garoffolo}, A., {Elias-Rosa}, N., {Benetti}, S., {et~al.} 2014, MNRAS,
  445, 1647

\bibitem[{{Nicholl} {et~al.}(2013){Nicholl}, {Smartt}, {Jerkstrand}, {Inserra},
  {McCrum}, {Kotak}, {Fraser}, {Wright}, {Chen}, {Smith}, {Young}, {Sim},
  {Valenti}, {Howell}, {Bresolin}, {Kudritzki}, {Tonry}, {Huber}, {Rest},
  {Pastorello}, {Tomasella}, {Cappellaro}, {Benetti}, {Mattila}, {Kankare},
  {Kangas}, {Leloudas}, {Sollerman}, {Taddia}, {Berger}, {Chornock}, {Narayan},
  {Stubbs}, {Foley}, {Lunnan}, {Soderberg}, {Sanders}, {Milisavljevic},
  {Margutti}, {Kirshner}, {Elias-Rosa}, {Morales-Garoffolo}, {Taubenberger},
  {Botticella}, {Gezari}, {Urata}, {Rodney}, {Riess}, {Scolnic}, {Wood-Vasey},
  {Burgett}, {Chambers}, {Flewelling}, {Magnier}, {Kaiser}, {Metcalfe},
  {Morgan}, {Price}, {Sweeney}, \& {Waters}}]{Nicholl_etal_2013}
{Nicholl}, M., {Smartt}, S.~J., {Jerkstrand}, A., {et~al.} 2013, \nat, 502, 346

\bibitem[{{Ofek} {et~al.}(2013){Ofek}, {Fox}, {Cenko}, {Sullivan}, {Gnat},
  {Frail}, {Horesh}, {Corsi}, {Quimby}, {Gehrels}, {Kulkarni}, {Gal-Yam},
  {Nugent}, {Yaron}, {Filippenko}, {Kasliwal}, {Bildsten}, {Bloom},
  {Poznanski}, {Arcavi}, {Laher}, {Levitan}, {Sesar}, \&
  {Surace}}]{Ofek_etal_2013}
{Ofek}, E.~O., {Fox}, D., {Cenko}, S.~B., {et~al.} 2013, ApJ, 763, 42

\bibitem[{{Ofek} {et~al.}(2014){Ofek}, {Zoglauer}, {Boggs}, {Barri{\'e}re},
  {Reynolds}, {Fryer}, {Harrison}, {Cenko}, {Kulkarni}, {Gal-Yam}, {Arcavi},
  {Bellm}, {Bloom}, {Christensen}, {Craig}, {Even}, {Filippenko},
  {Grefenstette}, {Hailey}, {Laher}, {Madsen}, {Nakar}, {Nugent}, {Stern},
  {Sullivan}, {Surace}, \& {Zhang}}]{Ofek_etal_2014}
{Ofek}, E.~O., {Zoglauer}, A., {Boggs}, S.~E., {et~al.} 2014, ApJ, 781, 42

\bibitem[{{Panagia}(2003)}]{Panagia_2003}
{Panagia}, N. 2003, in Lecture Notes in Physics, Berlin Springer Verlag, Vol.
  598, Supernovae and Gamma-Ray Bursters, ed. K.~{Weiler}, 113--144

\bibitem[{{Pastorello} {et~al.}(2007){Pastorello}, {Smartt}, {Mattila},
  {Eldridge}, {Young}, {Itagaki}, {Yamaoka}, {Navasardyan}, {Valenti}, {Patat},
  {Agnoletto}, {Augusteijn}, {Benetti}, {Cappellaro}, {Boles}, {Bonnet-Bidaud},
  {Botticella}, {Bufano}, {Cao}, {Deng}, {Dennefeld}, {Elias-Rosa},
  {Harutyunyan}, {Keenan}, {Iijima}, {Lorenzi}, {Mazzali}, {Meng}, {Nakano},
  {Nielsen}, {Smoker}, {Stanishev}, {Turatto}, {Xu}, \&
  {Zampieri}}]{Pastorello_etal_2007_06jc}
{Pastorello}, A., {Smartt}, S.~J., {Mattila}, S., {et~al.} 2007, Nature, 447,
  829

\bibitem[{{Pastorello} {et~al.}(2008){Pastorello}, {Mattila}, {Zampieri},
  {Della Valle}, {Smartt}, {Valenti}, {Agnoletto}, {Benetti}, {Benn}, {Branch},
  {Cappellaro}, {Dennefeld}, {Eldridge}, {Gal-Yam}, {Harutyunyan}, {Hunter},
  {Kjeldsen}, {Lipkin}, {Mazzali}, {Milne}, {Navasardyan}, {Ofek}, {Pian},
  {Shemmer}, {Spiro}, {Stathakis}, {Taubenberger}, {Turatto}, \&
  {Yamaoka}}]{Pastorello_etal_2008_Ibn}
{Pastorello}, A., {Mattila}, S., {Zampieri}, L., {et~al.} 2008, MNRAS, 389, 113

\bibitem[{{Pastorello} {et~al.}(2010){Pastorello}, {Smartt}, {Botticella},
  {Maguire}, {Fraser}, {Smith}, {Kotak}, {Magill}, {Valenti}, {Young},
  {Gezari}, {Bresolin}, {Kudritzki}, {Howell}, {Rest}, {Metcalfe}, {Mattila},
  {Kankare}, {Huang}, {Urata}, {Burgett}, {Chambers}, {Dombeck}, {Flewelling},
  {Grav}, {Heasley}, {Hodapp}, {Kaiser}, {Luppino}, {Lupton}, {Magnier},
  {Monet}, {Morgan}, {Onaka}, {Price}, {Rhoads}, {Siegmund}, {Stubbs},
  {Sweeney}, {Tonry}, {Wainscoat}, {Waterson}, {Waters}, \&
  {Wynn-Williams}}]{Pastorello_etal_2010}
{Pastorello}, A., {Smartt}, S.~J., {Botticella}, M.~T., {et~al.} 2010, ApJL,
  724, L16

\bibitem[{{Pastorello} {et~al.}(2013){Pastorello}, {Cappellaro}, {Inserra},
  {Smartt}, {Pignata}, {Benetti}, {Valenti}, {Fraser}, {Tak{\'a}ts}, {Benitez},
  {Botticella}, {Brimacombe}, {Bufano}, {Cellier-Holzem}, {Costado}, {Cupani},
  {Curtis}, {Elias-Rosa}, {Ergon}, {Fynbo}, {Hambsch}, {Hamuy}, {Harutyunyan},
  {Ivarson}, {Kankare}, {Martin}, {Kotak}, {LaCluyze}, {Maguire}, {Mattila},
  {Maza}, {McCrum}, {Miluzio}, {Norgaard-Nielsen}, {Nysewander}, {Ochner},
  {Pan}, {Pumo}, {Reichart}, {Tan}, {Taubenberger}, {Tomasella}, {Turatto}, \&
  {Wright}}]{Pastorello_etal_2013}
{Pastorello}, A., {Cappellaro}, E., {Inserra}, C., {et~al.} 2013, ApJ, 767, 1

\bibitem[{{Pejcha} \& {Prieto}(2015)}]{Pejcha_Prieto_2015}
{Pejcha}, O., \& {Prieto}, J.~L. 2015, ApJ, 799, 215

\bibitem[{{Pooley}(2014)}]{Pooley_2014}
{Pooley}, D. 2014, in IAU Symposium, Vol. 296, IAU Symposium, ed. A.~{Ray} \&
  R.~A. {McCray}, 103--107

\bibitem[{{Pritchard} {et~al.}(2014){Pritchard}, {Roming}, {Brown}, {Bayless},
  \& {Frey}}]{Pritchard_etal_2014}
{Pritchard}, T.~A., {Roming}, P.~W.~A., {Brown}, P.~J., {Bayless}, A.~J., \&
  {Frey}, L.~H. 2014, ApJ, 787, 157

\bibitem[{{Pritchard} {et~al.}(2012){Pritchard}, {Roming}, {Brown}, {Kuin},
  {Bayless}, {Holland}, {Immler}, {Milne}, \& {Oates}}]{Pritchard_etal_2012}
{Pritchard}, T.~A., {Roming}, P.~W.~A., {Brown}, P.~J., {et~al.} 2012, ApJ,
  750, 128

\bibitem[{{Quimby} {et~al.}(2011){Quimby}, {Kulkarni}, {Kasliwal}, {Gal-Yam},
  {Arcavi}, {Sullivan}, {Nugent}, {Thomas}, {Howell}, {Nakar}, {Bildsten},
  {Theissen}, {Law}, {Dekany}, {Rahmer}, {Hale}, {Smith}, {Ofek}, {Zolkower},
  {Velur}, {Walters}, {Henning}, {Bui}, {McKenna}, {Poznanski}, {Cenko}, \&
  {Levitan}}]{Quimby_etal_2011}
{Quimby}, R.~M., {Kulkarni}, S.~R., {Kasliwal}, M.~M., {et~al.} 2011, Nature,
  474, 487

\bibitem[{{Roming} {et~al.}(2005){Roming}, {Kennedy}, {Mason}, {Nousek}, {Ahr},
  {Bingham}, {Broos}, {Carter}, {Hancock}, {Huckle}, {Hunsberger}, {Kawakami},
  {Killough}, {Koch}, {McLelland}, {Smith}, {Smith}, {Soto}, {Boyd},
  {Breeveld}, {Holland}, {Ivanushkina}, {Pryzby}, {Still}, \&
  {Stock}}]{Roming_etal_2005}
{Roming}, P.~W.~A., {Kennedy}, T.~E., {Mason}, K.~O., {et~al.} 2005, Space
  Science Reviews, 120, 95

\bibitem[{{Roming} {et~al.}(2009){Roming}, {Pritchard}, {Brown}, {Holland},
  {Immler}, {Stockdale}, {Weiler}, {Panagia}, {Van Dyk}, {Hoversten}, {Milne},
  {Oates}, {Russell}, \& {Vandrevala}}]{Roming_etal_2009_08ax}
{Roming}, P.~W.~A., {Pritchard}, T.~A., {Brown}, P.~J., {et~al.} 2009, ApJL,
  704, L118

\bibitem[{{Roming} {et~al.}(2012){Roming}, {Pritchard}, {Prieto}, {Kochanek},
  {Fryer}, {Davidson}, {Humphreys}, {Bayless}, {Beacom}, {Brown}, {Holland},
  {Immler}, {Kuin}, {Oates}, {Pogge}, {Pojmanski}, {Stoll}, {Shappee},
  {Stanek}, \& {Szczygiel}}]{Roming_etal_2012}
{Roming}, P.~W.~A., {Pritchard}, T.~A., {Prieto}, J.~L., {et~al.} 2012, arXiv:1202.4840

\bibitem[{{Russell} \& {Immler}(2012)}]{Russell_Immler_2012}
{Russell}, B.~R., \& {Immler}, S. 2012, ApJL, 748, L29

\bibitem[{{Smith} {et~al.}(2015){Smith}, {Mauerhan}, {Cenko}, {Kasliwal},
  {Silverman}, {Filippenko}, {Gal-Yam}, {Clubb}, {Graham}, {Leonard}, {Horst},
  {Williams}, {Andrews}, {Kulkarni}, {Nugent}, {Sullivan}, {Maguire}, {Xu}, \&
  {Ben-Ami}}]{Smith_etal_2015}
{Smith}, N., {Mauerhan}, J.~C., {Cenko}, S.~B., {et~al.} 2015, arXiv:1501.02820

\bibitem[{{Soderberg} {et~al.}(2008){Soderberg}, {Berger}, {Page}, {Schady},
  {Parrent}, {Pooley}, {Wang}, {Ofek}, {Cucchiara}, {Rau}, {Waxman}, {Simon},
  {Bock}, {Milne}, {Page}, {Barentine}, {Barthelmy}, {Beardmore}, {Bietenholz},
  {Brown}, {Burrows}, {Burrows}, {Byrngelson}, {Cenko}, {Chandra}, {Cummings},
  {Fox}, {Gal-Yam}, {Gehrels}, {Immler}, {Kasliwal}, {Kong}, {Krimm},
  {Kulkarni}, {Maccarone}, {M{\'e}sz{\'a}ros}, {Nakar}, {O'Brien}, {Overzier},
  {de Pasquale}, {Racusin}, {Rea}, \& {York}}]{Soderberg_etal_2008}
{Soderberg}, A.~M., {Berger}, E., {Page}, K.~L., {et~al.} 2008, Nature, 453,
  469

\bibitem[{{Stritzinger} {et~al.}(2012){Stritzinger}, {Taddia}, {Fransson},
  {Fox}, {Morrell}, {Phillips}, {Sollerman}, {Anderson}, {Boldt}, {Brown},
  {Campillay}, {Castellon}, {Contreras}, {Folatelli}, {Habergham}, {Hamuy},
  {Hjorth}, {James}, {Krzeminski}, {Mattila}, {Persson}, \&
  {Roth}}]{Stritzinger_etal_2012}
{Stritzinger}, M., {Taddia}, F., {Fransson}, C., {et~al.} 2012, ApJ, 756, 173

\bibitem[{{Thomas} {et~al.}(2011){Thomas}, {Aldering}, {Antilogus}, {Aragon},
  {Bailey}, {Baltay}, {Bongard}, {Buton}, {Canto}, {Childress}, {Chotard},
  {Copin}, {Fakhouri}, {Gangler}, {Hsiao}, {Kerschhaggl}, {Kowalski}, {Loken},
  {Nugent}, {Paech}, {Pain}, {Pecontal}, {Pereira}, {Perlmutter}, {Rabinowitz},
  {Rigault}, {Rubin}, {Runge}, {Scalzo}, {Smadja}, {Tao}, {Weaver}, {Wu},
  {(Nearby Supernova Factory)}, {Brown}, \& {Milne}}]{Thomas_etal_2011}
{Thomas}, R.~C., {Aldering}, G., {Antilogus}, P., {et~al.} 2011, ApJ, 743, 27

\bibitem[{{Thompson} {et~al.}(2004){Thompson}, {Chang}, \&
  {Quataert}}]{Thompson_etal_2004}
{Thompson}, T.~A., {Chang}, P., \& {Quataert}, E. 2004, ApJ, 611, 380

\bibitem[{{Walker} {et~al.}(2012){Walker}, {Hachinger}, {Mazzali}, {Ellis},
  {Sullivan}, {Gal Yam}, \& {Howell}}]{Walker_etal_2012}
{Walker}, E.~S., {Hachinger}, S., {Mazzali}, P.~A., {et~al.} 2012, MNRAS, 427,
  103

\bibitem[{{Wang} {et~al.}(2009){Wang}, {Li}, {Filippenko}, {Foley}, {Kirshner},
  {Modjaz}, {Bloom}, {Brown}, {Carter}, {Friedman}, {Gal-Yam}, {Ganeshalingam},
  {Hicken}, {Krisciunas}, {Milne}, {Silverman}, {Suntzeff}, {Wood-Vasey},
  {Cenko}, {Challis}, {Fox}, {Kirkman}, {Li}, {Li}, {Malkan}, {Moore},
  {Reitzel}, {Rich}, {Serduke}, {Shang}, {Steele}, {Swift}, {Tao}, {Wong}, \&
  {Zhang}}]{Wang_etal_2009_05cf}
{Wang}, X., {Li}, W., {Filippenko}, A.~V., {et~al.} 2009, ApJ, 697, 380

\bibitem[{{Woosley} \& {Bloom}(2006)}]{Woosley_Bloom_2006}
{Woosley}, S.~E., \& {Bloom}, J.~S. 2006, ARAA, 44, 507

\bibitem[{{Zeh} {et~al.}(2004){Zeh}, {Klose}, \& {Hartmann}}]{Zeh_etal_2004}
{Zeh}, A., {Klose}, S., \& {Hartmann}, D.~H. 2004, ApJ, 609, 952

\end{thebibliography}
\end{document}